\documentclass[12pt]{amsart}
\usepackage{pdfsync}
\usepackage{enumerate}
\usepackage{graphicx}
\usepackage{latexsym}
\usepackage{amsfonts}
\usepackage[utf8]{inputenc}
\usepackage{amsthm}
\usepackage{amsmath}
\usepackage{amssymb}
\usepackage[
            unicode,
            breaklinks=true,
            colorlinks=true]{hyperref}
\usepackage[active]{srcltx}
\usepackage{amscd}
\usepackage[all]{xy} 
\usepackage{mathrsfs} 
\usepackage{stmaryrd} 
\usepackage{amsopn} 
\usepackage{eepic}
\usepackage{epic}
\usepackage{eucal}
\usepackage{epsfig}
\usepackage{psfrag}

\newtheorem{theorem}{Theorem}

\newcommand{\Hilbert}{{\mathcal{H}}}  
\newcommand{\R}{{\mathbb R}}
\newcommand{\Z}{{\mathbb Z}}

\newcommand{\N}{{\mathbb N}}

\newcommand{\phy}{\varphi}

\newcommand{\op}[1]{\!\!\mathop{\rm ~#1}\nolimits}

\newcommand{\ii}{\mathrm{i}}

\newcommand{\om}{\omega}

\renewcommand{\leq}{\leqslant}

\newcommand{\abs}[1]{\left|#1\right|}
\newcommand{\OP}{{\rm Op}_\hbar}

\newenvironment{example}{\refstepcounter{theorem}\par\medskip\noindent{\bf
Example~\thetheorem.}}{\unskip\nobreak\hfill\hbox{ $\oslash$}\par\bigskip}

\newenvironment{definition}{\refstepcounter{theorem}\par\medskip\noindent{\bf
Definition~\thetheorem.}}{\unskip\nobreak\hfill\hbox{}\par\medskip}

\begin{document}

\title{Symplectic spectral geometry of semiclassical operators}

\author{{\'A}lvaro Pelayo}

\maketitle

\date{}

\begin{abstract}
In the past decade there has been a flurry of activity at the intersection of spectral theory
 and symplectic geometry. In this paper we review recent results on semiclassical spectral 
theory for commuting Berezin\--Toeplitz and $\hbar$\--pseudodifferential operators. 
The paper emphasizes the interplay between spectral theory of operators (quantum theory) 
and symplectic geometry of Hamiltonians (classical theory), with an eye towards recent
developments on the geometry of finite dimensional integrable systems. 
 \end{abstract}

\section{Introduction} \label{sec:intro}

This paper\footnote{Based  on a
 scheduled plenary Talk by the author at the 2012 Joint Congress of the
Belgian, Royal Spanish and Luxembourg Mathematical Societies.  I was unable to attend the
Congress and I thank the organizers for the invitation to write this paper.
This paper is \emph{not} a survey but rather a report on recent developments. Throughout we keep a more informal tone than
in a regular research paper. We refer to the articles \cite{PeVN2011b, PeVN2012} for 
details and further bibliographic references.}  gives a concise
exposition of some recent results on spectral theory of $\hbar$\--pseudodifferential
and Berezin\--Toeplitz operators. Most of what I will say is contained in  
my papers with   L.~Charles,  L.~Polterovich, and 
S.~V\~{u} Ng\d{o}c \cite{ChPeVN2011, PeVN2012,PePoVN2012}.  
I will also discuss some recent works on symplectic geometry of finite
dimensional completely integrable Hamiltonian systems by  the author and
V\~{u} Ng\d{o}c \cite{PeVN2009, PeVN2011} because they
are central to the spectral theory. These papers contain
classification results for the so called completely integrable Hamiltonian systems
of \emph{semitoric type}, and are 
in the spirit of the seminal papers of Atiyah \cite{At1982}, Guillemin\--Sternberg \cite{GuSt1982},
and Delzant \cite{d} on toric systems.

In this paper we are going to emphasize the connection of symplectic geometry 
with spectral theory and microlocal analysis (see Guillemin\--Sternberg~\cite{GuSt2012} 
and Zworski~\cite{Zw2012}). In fact, the development of semiclassical
microlocal analysis in the past four decades now allows a fruitful  
interplay between symplectic geometry (classical mechanics) 
and spectral theory (quantum mechanics). 
The literature on these subjects is vast and I refer to the aforementioned
papers for a more comprehensive list of references.

\section{Symplectic geometry and integrable systems}

The word ``symplectic" was introduced by H. Weyl (Elmshorn 1885-Z\"urich 1955)
in his book on classical groups \cite{We1939}. 
It derives from a Greek word meaning \emph{complex}. 
Symplectic geometry studies symplectic manifolds.
A \emph{symplectic manifold} is a pair $(M,\omega)$ consisting of
a smooth $\op{C}^{\infty}$\--manifold $M$ and a closed and non\--degenerate
differential $2$\--form $\omega$ on it, called a \emph{symplectic form}. 
For instance, we can take $M$ to be a surface, 
and $\omega$ to be an area form on it (in dimension $2$, a symplectic form is
the same as an area form).  Another typical example is $\mathbb{R}^{2n}$
equipped with coordinates $(x_1,y_1,\ldots,x_n,y_n)$ and symplectic
form 
$
\sum_{i=1}^n {\rm d} x_i \wedge {\rm d}y_i.
$ 
The cotangent bundle of any compact smooth manifold is also a symplectic
manifold in a natural way.

Symplectic manifolds are even\--dimensional (because 
the symplectic form is non\--degenerate) and orientable (because $\omega^{{\rm dim}\,M/2}$ is a volume form).
Let's write $2n$ for the dimension of $M$. 
If $M$ is  compact, then one can use Stokes' theorem to show that for every $0\leq k\leq n$ we have
that $0 \neq [\omega]^k \in\op{H}^{2k}_{\textup{dR}}(M).$ 
By a famous theorem of Darboux  \cite{darboux}, near each point in $(M,\, \omega)$ there exist
coordinates $( x_{1},y_{1},\ldots,x_n,y_n)$ such that 
$
\omega
$
in which $\omega$ has the form
$
\sum_{i=1}^n {\rm d} x_i \wedge {\rm d}y_i,
$ 
so symplectic manifolds have \emph{no} local invariants, except the dimension.

\begin{figure} [h]
  \centering
  \includegraphics[width=\textwidth]{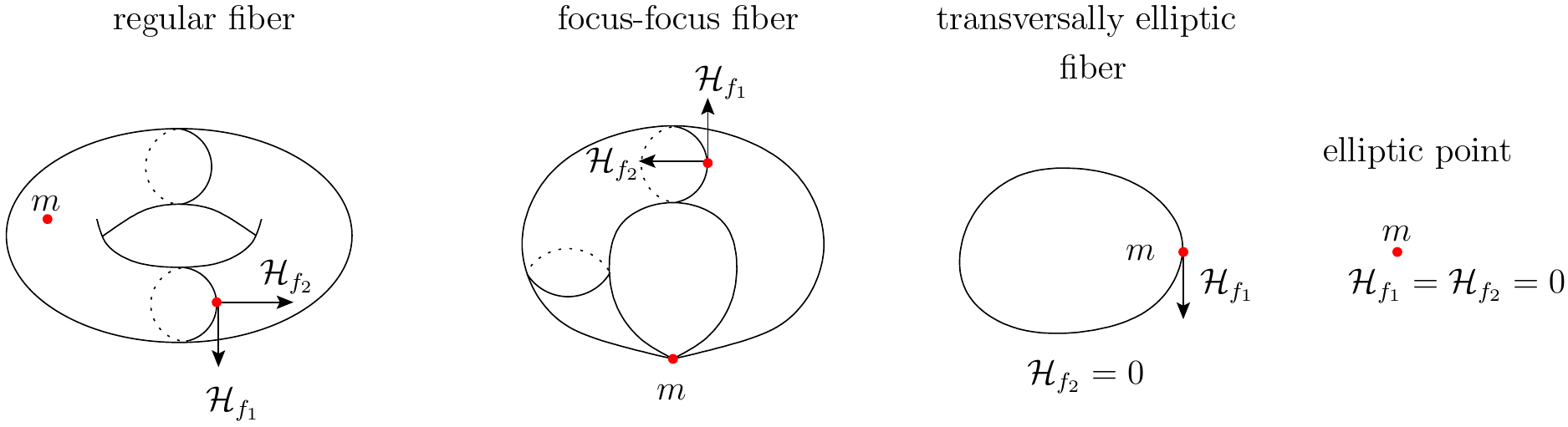}
  \caption{Some possible singularities of an
    integrable system.}
  \label{singularities}
\end{figure}

One important class of dynamical systems which 
can be studied with the tools of symplectic geometry 
are those called \emph{integrable}. 

\begin{definition}
A \emph{completely integrable system} (or simply an \emph{integrable system}) 
on a $2n$\--dimensional symplectic manifold 
$(M,\, \omega)$  is a smooth map
$
F:=(f_1,\, \ldots,\, f_n) \colon M \longrightarrow \mathbb{R}^n
$
such that each $f_i$ is constant along the flow\footnote{That is, any two $f_i,f_j$ 
commute in the sense that the Poisson brackets vanish:
$$
\{f_i, \, f_j\}:=\omega(\mathcal{H}_{f_i},\,\mathcal{H}_{f_j})=0,
\quad \textup{for all} \quad 1 \leq i,\,j \leq n.
$$} of each \emph{Hamiltonian vector field} 
$\mathcal{H}_{f_j}$,
where $\mathcal{H}_{f_j}$ is defined by Hamilton's equation 
$\omega(\mathcal{H}_{f_j},\,\cdot)=-{\rm d} f_j$
and, moreover,  the vector fields $\mathcal{H}_{f_1},\, \ldots,\, \mathcal{H}_{f_n}\,\,$ are linearly independent
 almost everywhere on $M$. 
 \end{definition}

 A \emph{singularity} is point $m \in M$ at which the vector fields
$\mathcal{H}_{f_1},\, \ldots, \,\mathcal{H}_{f_n}$ are
 linearly dependent.  
 There are many mechanical systems which are integrable, for instance: the \emph{coupled spin\--oscillator} (also
 called Jaynes\--Cummings model, see \cite{BaCaDo2009}), 
 the \emph{spherical pendulum}, the \emph{two\--body problem},  the 
\emph{Lagrange top}, or the \emph{Kowalevski top}.  All of these systems have singularities.

 While there are a few results on symplectic theory of
 integrable systems, the subject is largely not understood, we refer to \cite{PeVN2011b,PeVN2012} 
 for a more extensive discussion. In particular, \cite{PeVN2011b} aims to give a more comprehensive 
 description of the current state of the art of the symplectic theory of integrable systems. It is
 interesting to note that features about the symplectic geometry of singularities can be
 detected using spectral theory, see for instance \cite{CMP}, where this is done for some
 of the singularities of the coupled spin\--oscillator.

\section{Notions of spectrum}

\subsection{Classical and quantum spectra}
The self\--adjoint operators $T_1,\ldots,T_d$ on a Hilbert space are mutually commuting
if  their spectral
measures $\mu_1,\dots,\mu_d$ pairwise commute.
Then one can define the joint spectral measure on $\R^d$~:
$$
\mu := \mu_1\otimes\cdots\otimes\mu_d.
$$

\begin{definition}
The \emph{joint spectrum} of $(T_1,\ldots,T_d)$ is 
the support of the joint spectral measure. It is denoted by
$\op{JointSpec}(T_1,\, \ldots, \, T_d).$ 
\end{definition}

For instance, if the $T_j$'s are endomorphisms of a finite dimensional
vector space, then the joint spectrum of $T_1,\ldots,T_d$ is the set
$$
\Big\{(\lambda_1,\dots,\lambda_d)\in \mathbb{R}^d \,\, | \,\, \exists v \neq 0 \,\, \textup{such that}\,\,
P_j v = \lambda_j v \,\,\,\, \forall j=1,\dots,n\Big\}.
$$

If $T_1,\, \ldots,\, T_d$ are pairwise commuting semiclassical
operators, then of course the joint spectrum of
  $T_1,\ldots,T_d$ depends on the semiclassical parameter $\hbar$.

Following the physicists, we use the following definition.

\begin{definition}
We call \emph{classical
  spectrum} of $(T_1,\dots, T_d)$ the image $$F(M)\subset \R^d,$$ where
$F=(f_1,\dots,f_d)$ is the map of principal symbols of
$T_1,\dots,T_d$.
\end{definition}

\begin{figure}[htbp]
  \begin{center}
    \includegraphics[height=5cm, width=7cm]{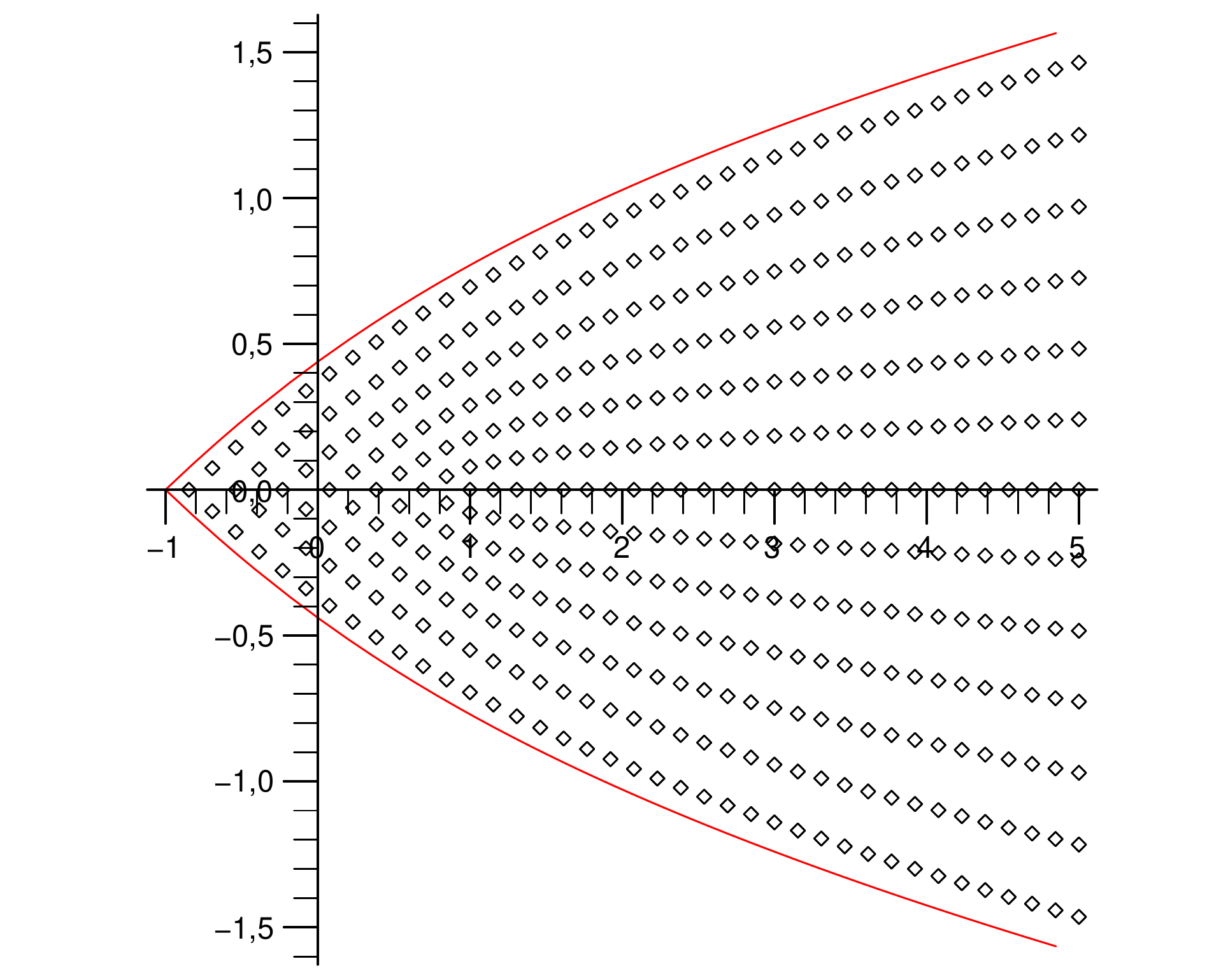}
    \caption{Joint spectrum of quantum \emph{Jaynes\--Cummings model}.}
    \label{fig:ff}
  \end{center}
\end{figure}

\subsection{Jaynes\--Cummings model}
An interesting system given by two self\--adjoint commuting operators
is the quantum \emph{Jaynes\--Cummings model}, studied in detail in \cite{PeVN2012}, and
which is given as follows. For any
  $\hbar>0$ such that $2=\hbar(n+1)$, for some non-negative integer
  $n\in\N$, let $\mathcal{H} \subset {\rm L}^2(\mathbb{R})$ denote the standard $n+1$-dimensional
  Hilbert space quantizing the sphere $S^2$. Consider the operators:
  \begin{eqnarray}
    \hat{x}:=\frac{\hbar}{2}(a_1a_2^*+a_2a_1^*), \qquad
    \hat{y}:=\frac{\hbar}{2\ii }(a_1a_2^*-a_2a_1^*), \qquad
    \hat{z}:=\frac{\hbar}{2}(a_1a_1^*-a_2a_2^*). \nonumber
  \end{eqnarray}
  where
$$
a_i:=\frac{1}{\sqrt{2 \hbar}} \Big( \hbar
\frac{\partial}{\partial x_j} +x_j\Big),\,\,\,\,i=1,2.
$$
 The operators on the Hilbert
   space $\mathcal{H} \otimes \op{L}^2(\R)\subset \op{L}^2(\R^2)
   \otimes \op{L}^2(\R)$ 
   $$
   \hat{f_1}:=\op{Id} \otimes \Big(
   -\frac{\hbar^2}{2} \frac{\op{d}^2}{\op{d}u^2} +\frac{u^2}{2} \Big)
   + (\hat{z} \otimes \op{Id})
   $$
   and
   $$
   \hat{f_2}:=\frac{1}{2}(\hat{x}\otimes u + \hat{y} \otimes
   (\frac{\hbar}{\ii}\frac{\partial}{\partial u})),
   $$
   are unbounded, self\--adjoint, and commute. The
   spectrum of $\hat{f_1}$ is discrete and consists of eigenvalues in
   $$\hbar\,\Big(\frac{1-n}{2}+\N\Big).$$ The joint spectrum for a fixed value of
   $\hbar$ is depicted in Figure \ref{fig:ff}. This quantum model  is in fact constructed 
   by hands on quantization of the classical system given by the symplectic
  manifold $M=S^2\times\R^2$, where $S^2$ is viewed as the unit
  sphere in $\R^3$ with coordinates $(x,\,y,\,z)$, and the second
  factor $\R^2$ is equipped with coordinates $(u,\, v)$, and
  the Hamiltonians  $f_1 := (u^2+v^2)/2 + z$ and $f_2 := \frac{1}{2} \, (ux+vy)$.
  So $f_1$ and $f_2$ are the principal symbols of $\hat{f_1}$ and $\hat{f}_2$.

 \section{Classical from semiclassical spectra} \label{sec:toeplitz}

\subsection{Compact case}
 Let $(M,\,\omega)$ be a compact
 symplectic manifold whose symplectic form represents an integral de
 Rham cohomology class of $M$.  In what follows such symplectic
 manifolds will be called {\it prequantizable}. They admit a
 prequantum line bundle $\mathcal{L}$. 
 Assume that $M$ is endowed with a
 complex structure $j$ compatible with $\om$, so that $M$ is \emph{K{\"a}hler}
 and $\mathcal{L}$ is \emph{holomorphic}. Here the holomorphic structure of
 the prequantum bundle is the unique one compatible with the
 connection.  
 
 For a positive integer
 $k=1/\hbar$,  we write $\mathcal{H}_{\hbar}$ for the space 
 $\mathrm{H}^0(M,\mathcal{L}^k)$ of holomorphic sections of $\mathcal{L}^k$. 
 Since $M$ is compact, $\mathcal{H}_{\hbar}$ is a closed finite
 dimensional subspace of the Hilbert space ${\rm L}^2(M,
 \mathcal{L}^k)$. Here the scalar product is defined by integrating
 the Hermitian pointwise scalar product of sections against the
 Liouville measure of $M$.  
 
 Denote by $\Pi_{\hbar}$ the orthogonal
 projector of ${\rm L}^2(M, \mathcal{L}^k)$ onto
 ${\mathcal{H}}_{\hbar}$. In this  case, we have the following definition.

 \begin{definition}
 A \emph{Berezin\--Toeplitz operator} is a sequence
$$T:=(T_{\hbar} \colon \Hilbert_{\hbar} \rightarrow \Hilbert_{\hbar})_{\hbar=1/k; \; k \in \mathbb{N}^*}$$ of operators of the form
$$
   (T_{\hbar} := \Pi_{\hbar} f(\cdot,\, k))_{k\in \mathbb{N}^*},
$$
where $f(\cdot,\,k)$, viewed as a multiplication operator, is a
 sequence in ${\rm C}^{\infty}(M)$ with an asymptotic expansion
 $$
 f_0 + k^{-1} f_1 + \ldots
 $$
 for the ${\rm C}^{\infty}$ topology. The coefficient $f_0$ is 
 the {\em principal symbol} of $(T_{\hbar})_{\hbar=1/k; k \in \mathbb{N}^*}$. 
 \end{definition}

Before stating the result of this section, recall that
the \emph{Hausdorff distance}  ${\rm d}_H(A,\,B)$ between two subsets $A$ and $B$
of $\R^n$ is the infimum of the $\epsilon > 0$ such that $A \subseteq
 B_\epsilon$ and $B \subseteq A_\epsilon$,
where for any subset $X$ of $\R^n$, the set $X_{\epsilon}$ is
$$
X_\epsilon := \bigcup_{x \in X} \{m \in \R^n\, \, | \,\, \|x - m \|
\leq \epsilon\}. 
$$
 If $(A_k)_{k \in \N^*}$ and $(B_k)_{k \in \N^*}$
are sequences of subsets of $\mathbb{R}^n$, we say that $A_k = B_k +
\mathcal{O}(k^{-\infty})$ if 
$${\rm
  d}_H(A_k,\,B_k)=\mathcal{O}(k^{-N})\,\,\,\,\, \forall N \in \N^*. 
  $$
  In the
  following theorem, the convergence is taken in the Hausdorff metric.

 \begin{figure}[h]
   \centering
   \includegraphics[width=0.53\textwidth]{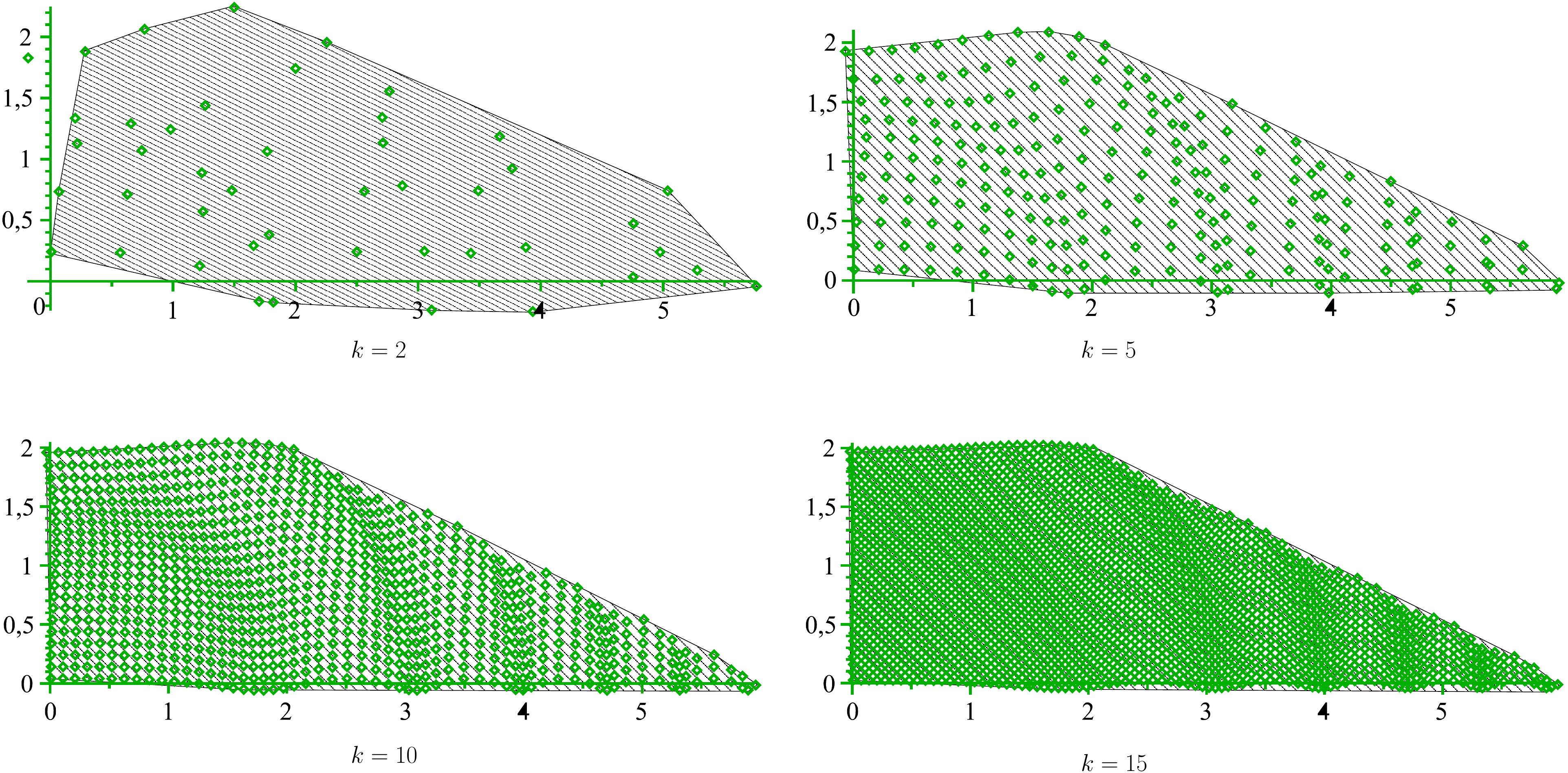}
   \caption{Convergence of convex hulls of spectra}
   \label{fig:convex_hull}
 \end{figure}

\begin{theorem}[Pelayo\--Polterovich\--V\~u Ng\d oc \cite{PePoVN2012}]  \label{theo:1}
Let 
  $\mathcal{F}_d:=(T_1,\dots, T_d)$ be a family of pairwise commuting self\--adjoint Berezin-Toeplitz operators
  on $M$.  Let $\mathcal{S}_d\subset\R^d$ be the classical spectrum of
  $\mathcal{F}_d$, and suppose that it is a convex set.  Then 
$
\textup{JointSpec}(\mathcal{F}_d) \to \mathcal{S}_d,
$
as $\hbar \rightarrow 0$.
\end{theorem}

The proof of Theorem~\ref{theo:1} uses microlocal techniques (the key lemma for the
proof is \cite[Lemma 5]{PePoVN2012}). 
The result proven in \cite{PePoVN2012} is stronger than Theorem \ref{theo:1}: one
does not need to assume that $\mathcal{S}_d$ is convex. In the general case, what we proved is that
we have convergence (still in the Hausdorff metric) at the level of convex hulls
$$
\textup{Convex Hull}(\textup{JointSpec}(\mathcal{F}_d)) \to \textup{Convex Hull}(\mathcal{S}_d),
$$
as $\hbar \to 0$ (see  Figure \ref{fig:convex_hull}). 
These results may be extended in a natural way to non\--commuting operators,
see \cite[Section 9]{PePoVN2012}.

 \subsection{Noncompact case} \label{sec:toeplitz}

\label{sec:pseudodifferential}

Suppose that $M$ is $\R^{2n}$, or the cotangent bundle ${\rm T}^*X$ of a compact smooth $n$\--dimensional 
manifold $X$  (with a smooth density $\mu$). In these cases a semiclassical quantization of $M$ is given by
semiclassical $\hbar$\--pseudodifferential operators, a well\--known semiclassical version
of the quantization given by homogeneous pseudodifferential operators (see for
instance Dimassi\--Sj\"ostrand~\cite{dimassi-sjostrand}). 
Symbolic calculus of pseudodifferential operators holds
when the symbols belong to a H\"omander class, eg. take
$\mathcal{A}_0$ consisting of functions $f \in \textup{C}^\infty(\R^{2n}_{(x,\xi)})$ such that
there exists $m\in\R$ for which $$\abs{\partial_{(x,\xi)}^\alpha f}\leq C_\alpha\langle
(x,\xi)\rangle^m$$ for all $\alpha\in\N^{2n}$.
Here $\langle z \rangle:= (1+|z|^2)^{1/2}$.

 If $f\in\mathcal{A}_0$, its \emph{Weyl quantization} is defined on
 $\mathcal{S}(\R^n)$ by
 \begin{equation}
   \label{equ:weyl}
   (\OP(f) u)(x) := \frac{1}{(2\pi\hbar)^{n}}\int_{\R^{2n}}
   \textup{e}^{\frac{\textup{i}}{\hbar}((x-y)\cdot\xi)} f({\textstyle\frac{x+y}{2}},\xi)u(y)\textup{d}y \textup{d}\xi.
   \nonumber
 \end{equation}
 
Let's cover $X$ with charts $U_1,\dots,U_N$, each of which is identified with a convex bounded  domain of $\R^n$ equipped with the Lebesgue measure.  Now consider a partition of unity
$\chi^2_1,\dots,\chi^2_n$ subordinated to $U_1,\dots,U_N$. Let
 $f\in\textup{C}^\infty(\textup{T}^*X)$ such that
$
\abs{\partial_\xi^\alpha f(x)} \leq C_\alpha \langle \xi \rangle ^m
$ for all $(x,\xi)\in \textup{T}^*X,\,\alpha\in \N^n$,
for some $m\in\R$. Let $\OP^j(f)$ be the \emph{Weyl quantization} calculated in $U_j$ and
define: 
\begin{equation} \label{eq-weyl-manifold}
\OP(f) u:= \sum_{j=1}^N \chi_j \cdot \OP^j(f) (\chi_j u)\;, \textup{for}\,\, u\in\textup{C}^\infty(X)\,, \nonumber
\end{equation}
which is a pseudodifferential operator on $X$ with principal symbol $f=\sum_{i=1}^N f\chi_j^2$.

In what follows we  work with the standard H\"{o}rmander class of symbols depending
on $\hbar$, $a(x,\xi,\hbar)$ on 
$\mathbb{R}^{2n}$ or ${\rm T}^*X$ with compact $X$. 
We say that $a$ {\it mildly depends on $\hbar$} if $$a(x,\xi,\hbar)=a_0(x,\xi) + \hbar\, a_{1,\hbar}(x,\xi),$$
where all $a_{1,\hbar}(x,\xi)$ are uniformly bounded in $\hbar$
 and supported in the same compact set.\footnote{Note: the principal symbol $a_0$ can be unbounded.} 

\begin{definition}
A \emph{semiclassical} $\hbar$\--\emph{pseudodifferential operator on} $X$ is any sequence 
of the form $T:=(\OP(f))_{\hbar \in (0,\,1]}.$ 
\end{definition}

The analysis of $\hbar$\--pseudodifferential operators is delicate due to 
the possible unboundedness of the operators.

\begin{theorem}[Pelayo\--Polterovich\--V\~u~Ng\d oc \cite{PePoVN2012}]
 \label{theo:pseudodifferential}  
Let $X$ be either $\mathbb{R}^n$, or a
  closed manifold. Let  $\mathcal{F}_d:=(T_1,\dots T_d)$ be
 a family of pairwise commuting self\--adjoint semiclassical $\hbar$-pseudodifferential operators
  on $X$ whose symbols mildly depend on $\hbar$.  Let $\mathcal{S}_d\subset\R^d$ be 
  the classical spectrum of
  $\mathcal{F}_d$, and suppose that it is a convex set.  Then from the family
$
\{ \textup{JointSpec} (\mathcal{F}_d)\}_{\hbar \in J}
$
one can recover $\mathcal{S}_d$.
If moreover each operator $T_i$ is bounded, $1 \leq i \leq d$, then
$
\textup{JointSpec}(\mathcal{F}_d) \to \mathcal{S}_d,
$
as $\hbar \rightarrow 0$.
\end{theorem}

As it was explained before, a more general result holds, where one does not
need to assume convexity of the classical spectrum.

\begin{example}
The results in this section apply to a number of examples.
For instance, Theorem \ref{theo:pseudodifferential} applies
to the system given by a particle in a rotationally symmetric
potential (\cite[Section~9.2]{PePoVN2012}). Theorem~\ref{theo:1} applies to
systems given by Hamiltonian torus actions (\cite[Section~8.2]{PePoVN2012}), 
and to the coupled\--angular momenta  system (\cite[Section~8.3]{PePoVN2012}).
\end{example}

\section{Spectral theory for systems of toric type}

\subsection{Symplectic geometry}

An $n$\--tuple of smooth functions
$$(\mu_1,\dots,\mu) \colon M\to\mathbb{R}^n$$ on a $2n$\--dimensional
symplectic manifold $(M,\omega)$ is called a \emph{momentum map} for a
Hamiltonian $n$\--torus action if the Hamiltonian flows $t_j\mapsto
\phy_{\mu_j}^{t_j}$ are periodic of period 1, and pairwise commute:
$$\phy_{\mu_j}^{t_j}\circ \phy_{\mu_i}^{t_i} = \phy_{\mu_i}^{t_i} \circ
\phy_{\mu_j}^{t_j}$$ so that they define an action of $\R^n/\Z^n$.
We say that $(M,\omega,\mu)$ is a \emph{toric integrable system}, or
simply a \emph{toric system}, if in addition $M$ is compact and connected,
and the action of $\R^n/\Z^n$ is effective.  

Two  toric systems $(M, \om ,\mu)$ and $( M',
\om' , \mu')$ are \emph{isomorphic} if there exists a symplectomorphism
$\varphi: M \rightarrow M'$ such that $$\varphi^* \mu' = \mu.$$  The convexity
theorem of Atiyah \cite{At1982} and Guillemin\--Sternberg \cite{GuSt1982}
implies that $\mu(M)$ is a convex polytope in $\mathbb{R}^n$. 
By the Delzant classification theorem \cite{d}, $\mu(M)$ is a so called
\emph{Delzant polytope} (i.e. rational, simple, and smooth), and 
the toric integrable system $(M,\omega,\mu)$ is classified, up to isomorphisms, by $\mu(M)$.

\subsection{Semiclassical spectral theory}

In the case of toric integrable systems, a complete description of the semiclassical
spectral theory can be given.  That is, we are going to make a much stronger assumption than in 
 Section \ref{sec:toeplitz}: ``being toric"; but we are also going to obtain much more information 
 (in fact, \emph{all} the information). Let 
$(\mu_1,\ldots,\mu_n)$ be a toric integrable system on a compact prequantizable 
symplectic manifold $M$
 equipped with a prequantum bundle ${\mathcal{L}}$ and a
  compatible complex structure $j$.

\begin{theorem}[Charles\--Pelayo\--V\~{u}~Ng\d{o}c \cite{ChPeVN2011}] \label{theo:spectral} 
Let $T_1,\dots, T_n$ be commuting  Berezein\--Toeplitz operators  with
principal symbols $\mu_1,\ldots,\mu_n$. Then $$\textup{JointSpec}(T_1,\dots, T_n)$$
  is given by
   $$
  g \Big( \mu(M)\cap \biggl( v +
  {\tiny \frac{2 \pi}{k}} \mathbb{Z}^n \biggr) ;\, k \Big) +
  \mathcal{O}(k^{-\infty}),
  $$
  where $v$ is any vertex of $\mu(M)$ and
  $g(\cdot;k):\mathbb{R}^n\to\mathbb{R}^n$ admits a ${\rm C}^{\infty}$\--asymptotic expansion of
  the form
  $
  g(\cdot;k) = \textup{Id}+k^{-1}g_1+k^{-2}g_2+\cdots
  $
  where each $g_j:\mathbb{R}^n\to\mathbb{R}^n$ is smooth. 
  \end{theorem}

 Moreover, the multiplicity of the eigenvalues in Theorem \ref{theo:spectral} can be
described:  for all sufficiently large $k$, the multiplicity of the eigenvalues of 
 $\textup{JointSpec}(T_1,\dots, T_n)$ 
  is $1$, and there exists a small constant $\delta>0$ such that each ball of radius $\frac{\delta}{k}$ centered at  an eigenvalue contains precisely
  only that eigenvalue.

\subsection{Semiclassical isospectrality}

The following  type of inverse conjecture is classical and belongs to the realm
of questions in inverse spectral theory, going back to similar
questions raised (and in many cases answered) by pioneer works of
Colin de Verdi{\`e}re \cite{CdV, CdV2}.
It follows from combining the Delzant theorem with Theorem \ref{theo:spectral}.
Let $(M, \, \omega, \, \mu : M \rightarrow \R^n)$ be a toric integrable system 
  with a prequantum bundle ${\mathcal{L}}$ and a compatible complex structure $j$.

\begin{theorem}[Charles\--Pelayo\--V\~{u}~Ng\d{o}c \cite{ChPeVN2011}] \label{theo:inverse-spectral}
  Let $\mathcal{F}_n:=(T_1,\dots, T_n)$ be a family of commuting self\--adjoint
  Berezin\--Toeplitz operators  with principal symbols $\mu_1,\ldots,\mu_n$. Then one can
recover $(M,\omega,\mu)$ from the limit of the joint spectrum of $T_1,\ldots,T_n$.
\end{theorem}

In fact, combining Delzant's theorem with Theorem \ref{theo:1} gives an easier
proof of Theorem \ref{theo:inverse-spectral} which does not use Theorem \ref{theo:spectral}
which is a more difficult (but much more informative) result.

\section{Spectral theory for  systems of semitoric type}

\subsection{Symplectic geometry}

 A \emph{semitoric system} consists of a connected symplectic
four\--dimensional manifold $(M,\, \omega)$ and two smooth functions
$f_1 \colon M \to \R$ and $f_2 \colon M \to \R$ such that 
$f_1$ is constant along the flow of the Hamiltonian vector
  field $\mathcal{H}_{f_2}$ generated by $f_2$ or, equivalently,
  $\{f_1,\,f_2\}=0$ and for almost all points $p \in M$, the vectors
  $\mathcal{H}_{f_1}(p)$ and $\mathcal{H}_{f_2}(p)$ are linearly independent.
  Moreover,\footnote{this is the condition which gives rise to the name
  ``semitoric"}  $f_1$  is the
  momentum map of an $S^1$\--action on $M$, and it is a proper map.
  Finally, we require   $F:=(f_1,f_2) \colon M \to \mathbb{R}^2$ to have only non\--degenerate singularities without
  hyperbolic components.     Two semitoric systems 
  $$(M_1,\omega_1,F_1:=(f^1_1,f^1_1)) \,\, \textup{and}\,\, (M_2,\omega_2,F_2:=(f^2_1,f^2_2)) 
  $$
  are
  \emph{isomorphic} if there exists a symplectomorphism $\phi \colon
  M_1\to M_2$, and a smooth map $\varphi \colon F_1(M_1)\to \mathbb{R}$ with
  $\partial_2 \varphi \neq 0$, such that 
  \[
\begin{cases}
\phi^*f^1_1=  f^2_1 \\
\phi^*f_2 =
  \varphi(f^1_1,f^1_2).
\end{cases}
\]  

Semitoric systems can be classified, up to isomorphisms, in terms of
five symplectic invariants.  This classification appeared in \cite{PeVN2009,PeVN2011}.
Roughly speaking, these invariants are as follows: an integer $m_f$
  counting the number of isolated singularities, a collection of
Taylor series classifying
  symplectically a saturated neighborhood of the singular fiber corresponding
  to these sigularities, a  family of rational convex polygons
$$
\left(\Delta,\,(\ell_j)_{j=1}^{m_f},\,(\epsilon_j)_{j=1}^{m_f}\right),
$$
which is constructed from the image $F(M) \subset \mathbb{R}^n$ 
of the system by performing a very
precise ``cutting" (the $\ell_j$'s are vertical lines cutting $\Delta$ with orientations $\epsilon_j=\pm1$), 
an invariant measuring the volumes of certain
submanifolds meeting at each of the isolated singularities, and, finally, 
a collection of integers
  measuring how twisted the Lagrangian fibration of the system is around the
  singularities.  

Toric systems are a particular case
of semitoric systems. 
If the system is toric, then four of the invariants do not appear (and the
remaining one is simpler: a polygon, instead of a class of polygons).

  \subsection{Semiclassical spectral theory}

The semiclassical spectral theory of semitoric systems is not yet
understood. In \cite[Section 9]{PeVN2012} it is conjectured that from the semiclassical joint spectrum of two 
self\--adjoint commuting operators one can recover the integrable
  system given by the principal symbols, up to symplectic
   isomorphisms, provided these principal symbols form a semitoric system.
A sketch of proof of this conjecture appeared in \cite[Section 3.2]{PeVN2012}.

\vspace{2mm}

\emph{Acknowledgements}.  The author
was partly supported by grants NSF DMS-0635607,
NSF CAREER DMS-1055897 and 
Spain Ministry of Science grant Sev-2011-0087.    He is grateful to  L.~Polterovich and
and S.~V\~{u} Ng\d{o}c for comments on a preliminary version of this paper. 
He also thanks L.~Charles, H.~Hofer, L.~Polterovich, T. Ratiu,
and S.~V\~{u} Ng\d{o}c for stimulating discussions.

\bibliographystyle{new} 

\noindent
\\
{\bf {\'A}lvaro Pelayo}\\
School of Mathematics, Institute for Advanced Study\\
Einstein Drive\\
Princeton, NJ 08540 USA.
\\
\\
\noindent
Washington University, Mathematics Department \\
One Brookings Drive, Campus Box 1146\\
St Louis, MO 63130-4899, USA.\\
{\em E\--mail}: \texttt{apelayo@math.wustl.edu},\, \, \texttt{apelayo@math.ias.edu} \\
{\em Website}: \url{http://www.math.wustl.edu/~apelayo/}\\

\end{document}